\documentclass{elsart3}
\usepackage{natbib}
\usepackage{epsfig}
\begin{document}
\begin{frontmatter}
\title{The Vertex Tracker at Future $e^+e^-$ Linear Colliders}

\author{M.~Battaglia}
\address{University of California, Berkeley CA, USA and\\
CERN, CH-1211 Geneva 23, Switzerland}
%==========================================================================
%
%==========================================================================

\begin{abstract}
The physics program of high energy $e^+e^-$ linear colliders relies on the 
accurate identification of fermions to study in details the profile of the 
Higgs boson, search for new particles and later probe the multi-TeV mass region 
by direct searches and precision electro-weak measurements. This paper reviews 
the requirements, conceptual design and sensor R\&D for the Vertex Tracker.
\end{abstract}
\end{frontmatter}
\vspace*{-1.0cm}
\section{Introduction}

\vspace{-0.5cm}

The LHC collider at CERN represents the next step in the study of the high energy 
frontier in particle physics. We expect that the data collected in $pp$ collisions 
will provide evidence of the Higgs bosons and of Supersymmetry, or of other New Physics 
signals. The Higgs bosons observation will be a decisive breakthrough 
in our understanding of the origin of mass and electro-weak symmetry breaking. 
Signals of New Physics will clarify the solution to the hierarchy problem of the 
Standard Model. But neither the precision study of the Higgs profile nor the investigation 
of New Physics will be completed at the LHC. There are measurements which will be limited 
in accuracy, while others may not be feasible at all.

A high energy, high luminosity linear collider (LC), able to deliver $e^+e^-$ collisions 
at center-of-mass energies $\sqrt{s} =$ 0.3-0.5~TeV with luminosity in excess to 
$10^{34}$~cm$^{-2}$s$^{-1}$, later upgradeable to about 1~TeV, is considered as the 
next large scale project in accelerator particle physics. Present projects focus on 
warm RF (mostly X-band as for the NLC and JLC projects) or super-conducting cavities 
(proposed by the TESLA collaboration) to achieve the needed gradients~\cite{Loew:se}.
Beyond it, multi-TeV collisions appear to be achievable at a linear collider, using a 
novel two-beam acceleration scheme at high frequency, developed in the CLIC 
study~\cite{clic}. 

The Vertex Tracker is expected to provide the jet flavor identification 
capabilities and the accurate event reconstruction that make the linear collider 
unique and allow its physics program.

If the Higgs boson exists and it is light, as the present data indicates, its 
couplings to fermions of different flavor, and hence of different mass, must be 
accurately measured as a fundamental proof of the Higgs mechanism of mass generation, 
as well as its self-coupling. 
Efficient flavor tagging in multi-jet events and determination of heavy quark charge will 
be instrumental to study signals of New Physics both through the direct production of new 
heavy particles, coupled predominantly to $b$ and $t$ quarks, and by precise electro-weak 
data at the high energy frontier.
Physics requirements push the vertex tracker specifications to new levels. While much has 
been learned in two decades of R\&D on Si detectors for the LHC experiments, the LC 
motivates today new and complementary directions. Its experimental environment, with its 
lower event rate and radiation flux, admits Si sensors that are substantially thinner, 
more precise and more segmented than at the LHC. Technologies which have not been 
applicable in the high radiation environment of proton colliders are also available, as 
well as sensors of new concept. But significant R\&D is required. CCD vertex detectors 
have already demonstrated very high resolution and segmentation with low
multiple scattering at SLD~\cite{Abe}. 
But for the CCD technology to be applicable to the LC improved 
radiation hardness and a factor 100-1000 increase in readout speed are required. 
Technologies successfully developed for the LHC program, such as hybrid pixel sensors, are 
sufficiently radiation hard and can be read out rapidly. But they now need to be developed 
into much thinner devices with smaller cell size to improve their position resolution. 
Finally new technologies, such as MAPS, SOI and DEPFET sensors, are emerging as other 
potentially attractive solutions. But they need to be demonstrated on large scales and be 
tailored to the LC applications. These developments need to be guided by a continued program 
of physics studies and detailed simulations to define the optimal design and technology 
choices.

Several Vertex Tracker designs have been proposed, relying on different sensor
technologies. They all share the use of pixel devices, due to the high particle density which 
disallows the use of microstrip detectors. Emphasis is placed on minimizing the material budget, 
to improve track extrapolation performances also at small momenta in multi-jet final states. 

\section{Experimental Conditions}

The Vertex Tracker at the LC will be exposed to background and radiation levels and to 
track densities unprecedented for $e^+e^-$ collider machines, though still lower 
compared to proton colliders. The main source of background in the interaction region is 
due to $e^+e^-$ pairs produced and bent in the intense electro-magnetic interaction of 
the colliding beams. Such pairs set the most stringent constraint on the Vertex Tracker 
geometry. The radius and maximum length of the innermost sensitive layer are defined by 
the envelope of the deflected pairs. The radial and longitudinal position of the point of 
crossing of the pair envelope can be approximated as 
function of the number of particles in a bunch, $N$, the solenoidal magnetic field, $B$, 
and the bunch length, $\sigma_z$ by:
\begin{eqnarray}
R [cm] = 0.35 \sqrt{\frac{N}{10^{10}} \frac{1}{B [Tesla]} z [cm] 
\frac {1}{\sigma_z [mm]}}
\label{eq:r}
\end{eqnarray}
\begin{eqnarray}
z [cm] = 8.3~ R^2 B [Tesla] \sigma_z [mm] \frac{10^{10}}{N} .
\label{eq:z}
\end{eqnarray}
Warm RF technology requires $\sigma_z$ to be small, while the field strength $B$ is 
limited by the optic and quadrupole requirements at the final focus. The inward bound on 
the detector radius is thus set at $\simeq$~1.5~cm, up to 1~TeV. This radius appears safe 
also for the collimation of syncrotron radiation. At a multi-TeV collider the innermost 
radius must be pushed to $\simeq$~3~cm.
%\begin{table}
%\begin{center}
%\begin{tabular}{|l|c|c|r|r|r|}
%\hline
%LC & $\sqrt{s}$ (GeV) & R (cm) & BX$^{-1}$ & 25 ns$^{-1}$ & train$^{-1}$\\
%\hline
%NLC   & ~500 & 1.2 & 0.100  & 1.80 & 9.5 \\
%TESLA & ~800 & 1.5 & 0.050  & 0.05 & 225.0 \\
%CLIC  & 3000 & 3.0 & 0.005  & 0.18 & 0.8 \\
%\hline
%\end{tabular}
%\end{center}
%\end{table}

Particle tracks in highly collimated jets also significantly contribute to the local track 
density in physics events. This is expected to be 0.2-1.0~hits~mm$^{-2}$ at 500~GeV, to 
reach 0.5-2.5~hits~mm${^-2}$ at 3.0~TeV. These figures are comparable to, or even exceed, 
those expected at the LHC: 0.03 hits~mm$^{-2}$ for proton collisions in ATLAS and 
0.9 hits~mm$^{-2}$ for heavy ion collisions in ALICE.
The dose due to charged particles is expected to be manageable: $\simeq$~50~krad~y$^{-1}$.
On the contrary, the neutron background may be important for the sensor technology choice. 
Neutrons are produced in electromagnetic interactions of the spent beams and radiated 
particles. The resulting fluxes are in principle large. However, they are reduced by 
several orders of magnitude by the masks at the vertex tracker position. Still, the 
estimated fluxes reaching the Si detectors are expected of order 
$10^9$ $n$~(1~MeV)~cm$^{-2}$~y$^{-1}$ for TESLA~\cite{Wagner:2001qw} and NLC and about one 
order of magnitude larger for CLIC at 3~TeV~\cite{clicrep}.

\section{Vertex Tracker Conceptual Design}

The Vertex Tracker will likely consist of a multi-layered barrel section, directly 
surrounding the beam-pipe, complemented by forward disks to ensure tracking down to 
small angles. Five layers should ensure standalone pattern recognition and tracking 
capabilities as well as redundancy. 

The strongest requirements on the impact parameter resolution are set by the need of 
efficiently disentangling $H^0 \to b \bar{b}$ from $H^0 \to c \bar{c}$ Higgs boson 
decays~\cite{Battaglia:2000jb}. This can be best done by exploiting the difference in 
invariant mass and multiplicity of the decay products. But for this method to be efficient, 
secondary particle tracks need to be identified by their significantly large impact 
parameter down to low momenta. The charm jet tagging efficiency degrades by a factor 
1.5-2.0, at constant purity, if the impact parameter resolution $\sigma_{ip}$ changes from 
5~$\mu$m~$\oplus$~5~$\mu$m/$p_t$ to 10~$\mu$m~$\oplus$~30~$\mu$m/$p_t$. Since jets are 
tagged in pairs, such loss corresponds to 2 to 4~times the equivalent data statistics.
Several other physics processes support these requirements. 
A multi-layered vertex tracker with the first sensitive layer at 1.5~cm from the 
interaction region and 1\%~$X_0$ of total thickness can provide the target 
$\sigma_{ip}=$5~$\mu$m~$\oplus$~5~$\mu$m/$p_t$. Single point resolution of 5~$\mu$m, 
or better, has been achieved with different techniques. 

The main challenge comes from the 
limit on the material budget. Several solutions are being studied ranging from 20~$\mu$m 
thick CCD ladders (0.06\%~$X_0$/layer) supported only at their ends to back-thinned hybrid 
pixel sensors (0.3\%~$X_0$/layer). 
Extracting the heat dissipated by the sensors and their electronics is 
another important issue in the engineering design of the vertex tracker and the material 
budget may be driven by the power dissipation. The typical value is of order of 
15~$\mu$W/pixel for CCDs, 40~$\mu$W/channel for HPS and 4~$\mu$W/pixel for MAPS. 
The total power for CCD sensors may be lowered to about 10~W if 1~V clocks are feasible. 
The additional dissipation from the driver and read-out electronics is less critical being 
confined outside the sensitive part of the detector. In addition CCDs may need to operate 
at low temperature to improve their radiation tolerance. The heat management may also 
depend on the bunch structure of the collider and will need to be studied in details. 
In particular pulsed power operation is being considered to profit of the collider low duty 
cycle and tests have started.

Finally suppression of noise and RF pick-up is essential, due to the large number of channels.

\section{Si Sensor Technology and R\&D}

\subsection{Charge Coupled Devices}

CCD sensors have characteristics which match in principle the main LC 
performance requirements. Their pixel size is small, giving single point 
resolution better than 4~$\mu$m, and the sensors are thin, $\simeq$~20~$\mu$m. 
Two main limitations remain: the read-out timing and the neutron radiation damage.
At a collider with the TESLA bunch structure, the $\simeq$~3~M pixels in the first 
layer, need to be read-out in not more than 50~$\mu$s to ensure a background hit 
density below 5~mm$^{-2}$. Therefore a read-out clock of about 50~MHz is necessary.
A novel column parallel read-out (CPCCD) scheme is being developed by the LCFI 
Collaboration~\cite{Stefanov:sw}. Prototypes have been designed and produced and 
are presently being tested, which operate with low-voltage clock amplitudes to reduce 
power dissipation. 
The most important radiation damage in CCDs is bulk Si displacements caused 
by heavy particles causing charge carrier trapping. Deep level bound states have 
lifetime longer than the inter-pixel transfer time and the charge is lost resulting 
in a drop of the charge transfer efficiency (CTE). This becomes particularly important 
since charges need to be transported over lengths of order of cm. Two possible techniques 
to improve the CTE are being studied: cooling the detector to increase the trapping 
lifetime and keep the trapping centers filled and filling traps with light pulse flushing to
avoid further charge loss. Tests have been performed~\cite{Brau:pt} and indicate 
that the signal loss can be lowered with light pulses. First results give a signal loss 
reduction from 29\% to 18\% after integrating $6.5 \times 10^9$~$n$~cm$^{-2}$ and can be 
improved with an optimsed setup.

\subsection{Hybrid Pixel Sensors}

Hybrid pixel sensors (HPS) have provided a reliable solution to 3D tracking from
LEP~2 to LHC. Their main limitations, due to the total sensor plus chip thickness 
and the limited single point resolution, may be overcome with a dedicated R\&D program. 
Beside the vertex tracker, HPS detectors offer a suitable technology also for 
forward tracking with good resolution and fast time-stamping. A scheme with interleaved 
nodes, extending that usefully applied to microstrip detectors, was proposed to improve 
the point accuracy by interpolating the charge sharing on neighboring read-out nodes. 
Test structures have been produced and successfully tested, providing a 
proof of principle~\cite{Battaglia:2000kc}. A single point resolution of $\simeq 3$~$\mu$m 
can be achieved, if tracks are sufficiently isolated. Now a dedicated R\&D program on 
back-thinning and bump-bonding represents the main focus, to reduce the detector thickness.

\subsection{MAPS Sensors}

Monolithic Active Pixel Sensors (MAPS) exploit the epitaxial layer of the CMOS wafer 
as detector substrate, to integrate the detector and the front-end readout electronics on 
the same silicon wafer, using standard VLSI CMOS technology. The development of MAPS 
detectors started with application as photon detectors where they are becoming 
increasingly popular. Their application to detection of m.i.p. signals was initiated as a 
LC R\&D~\cite{Claus:he,Turchetta:sx} and the first vertex tracker based on this technology 
is under construction for the STAR detector upgrade at RHIC~\cite{Matis:2002jv}. The signal 
is collected from the undepleted bulk or epitaxial layer where the charge carriers spread by 
thermal diffusion. Small pixel size and integrated electronics offer a good solution to the 
problems of resolution and layer thickness. Detector have been tested on particle beams and 
after irradiation. Tolerance to neutron fluxes has been established up to 
10$^{12}$ $n$/cm$^2$, which is well beyond the LC requirements. A full scale 1~M pixel 
sensors has proved that MAPS offer full efficient detectors with 2~$\mu$m accuracy and 
excellent two-track resolution. New developments are addressing the read-out speed and 
providing increased functionality, including data sparsification and integrated correlated 
double sampling. Test structures in 0.35~$\mu$m technology with 5~MHz column parallel 
readout are being evaluated.

\subsection{Other Options}

The variety of technologies for applications at the linear collider is further enriched 
by new concepts currently being investigated.

Another route toward monolithic sensors is the realization of FET devices integrated in 
high-resistivity fully depleted $n$ bulk which amplify the charge at the point of collection, 
avoiding losses. This scheme, adopted by DEPFET devices, provides full bulk sensitivity and the 
low input capacitance ensures low noise and have robust correlated double sampling capabilities. 
DEPFET sensors have been developed primarily for X-ray imaging. A dedicated R\&D for the LC vertex 
tracker has started~\cite{Richter:dn}.

Another attractive architecture for a monolithic pixel sensor is Silicon on insulator (SOI), 
where a Si film sits on a thin insulator over a high resistivity Si substrate acting as 
detecting volume. The read-out is built in the thin layer. There are a number of technological 
issues to be addressed in matching the pixel manufacturing technique with the CMOS processing.
SOI test structures have been fabricated 0.8 AMS technology~\cite{Amati:dq} and characterized. 
Recently signals from ionizing particles have been recorded, providing a first proof of 
principle of this design.

Emerging ion etching technologies have enabled the development of a new 3D detector 
scheme~\cite{Parker:1996ge}. In these detectors small diameter holes are drilled 
through the silicon wafer. Carriers drift perpendicular to the wafer thickness and 
normal to the particle trajectory. 3D sensors are characterized by good radiation 
tolerance and very fast time response, owing to their geometry. This makes them 
interesting for applications at small radius in the forward region.

In a farther future, the deposition of hydrogenated amorphous Si layer on ASIC may also 
become a competitive technology, bringing advantages both in terms of fast signals and, 
possibly, productions costs~\cite{Jarron:2003}.

\section{Conclusions}

An active and diversified R\&D program on Si sensors for LC applications is 
presently ongoing world-wide. It addresses issues which are complementary to the 
developments tailored to the LHC, while other aspects of detector engineering, services 
and read-out electronics will largely profits from the LHC experience. At present several 
detector architectures appear promising. However, it will be important to extend the 
R\&D phase, until the time of project approval and final detector design. 
As pixel sensors have a wide, interdisciplinary field of applications, ranging from 
structural biology to medical imaging and astrophysics, the linear collider R\&D effort 
is also significantly nested to those broader developments.

\end{document}